\documentclass[a4paper,twocolumn, superscriptaddress]{revtex4}

\usepackage{hyperref} 
\usepackage{graphicx}

\begin{document}

\title{Structure analysis of two-dimensional nonlinear self-trapped photonic lattices in anisotropic photorefractive media}

\author{Bernd Terhalle}
\affiliation{Institut f\"{u}r Angewandte Physik, Westf\"{a}lische Wilhelms-Universit\"{a}t, 48149 M\"{u}nster, Germany}
\author{Denis Tr\"{a}ger}
\affiliation{Institut f\"{u}r Angewandte Physik, Westf\"{a}lische Wilhelms-Universit\"{a}t, 48149 M\"{u}nster, Germany}
\author{Liqin Tang}
\affiliation{Institut f\"{u}r Angewandte Physik, Westf\"{a}lische Wilhelms-Universit\"{a}t, 48149 M\"{u}nster, Germany}
\affiliation{TEDA Applied Physical School, Nankai University, Tianjin 300457, China}
\author{J\"{o}rg Imbrock}
\affiliation{Institut f\"{u}r Angewandte Physik, Westf\"{a}lische Wilhelms-Universit\"{a}t, 48149 M\"{u}nster, Germany}
\author{Cornelia Denz}
\affiliation{Institut f\"{u}r Angewandte Physik, Westf\"{a}lische Wilhelms-Universit\"{a}t, 48149 M\"{u}nster, Germany}

\begin{abstract}
We generate experimentally different types of two-dimensional
self-trapped photonic lattices in a photorefractive medium and
analyze the induced refractive index change using two different
methods. One method gives the first experimental Fourier space analysis of
both linear and nonlinear self-trapped photonic lattices with periodic phase modulation 
using partially spatially incoherent multi-band excitation of the
lattice modes. The other method utilizes the waveguiding properties of
the lattice to achieve a real space analysis of the induced refractive
index change. The results of both methods are compared.

 \end{abstract}

\maketitle
\section{Introduction}
Periodic photonic structures attracted strong interest several
years ago, due to many novel possibilities for controlling light
propagation, beam steering and trapping. The idea behind this approach
is that a periodically modulated refractive index modifies the
diffraction relation and splits it into regions of allowed propagation
separated by forbidden band gaps \cite{Kivshar}. The concept is very similar to
solid state physics, where the periodic potential of the atoms leads
to the formation of band gaps for the propagation of electrons with a  wide variety of well-known applications, too.

Photonic lattices can be induced optically in photorefractive crystals by linear diffraction-free light patterns created by the interference of several plane waves \cite{Fleischer}.
However, the induced change of the refractive index depends on the light intensity and, in the nonlinear regime, is accompanied by the self action effect \cite{Chen}.
Another important possibility to create stationary two-dimensional light patterns for all optical induction is offered by self-trapped periodic waves. Here,
the diffraction-free light patterns in the form of stationary nonlinear periodic waves can propagate without change in their profile, becoming the eigenmodes of the self-induced periodic potentials.
Photonic lattices created by two-dimensional arrays of pixellike
solitons were demonstrated experimentally in parametric processes
\cite{Minardi} and in photorefractive crystals with both coherent
\cite{Petter} and partially incoherent \cite{Chen, Neshev, Martin}
light. For the case of two-dimensional arrays of in-phase solitons
created by amplitude modulation, every pixel of the lattice induces a
waveguide that can be manipulated by an external steering beam
\cite{Petter, Martin, Petrovic}. However, the spatial periodicity of
these arrays is limited. When the lattice constant gets too small,
the nonlocality of the induced refractive index change leads to
interactions between neighboring solitons and these interactions result
in a  strong instability of in-phase soliton arrays. In contrast, the recently demonstrated two-dimensional lattices of  out-of-phase solitons can be made robust with smaller lattice spacing \cite{Desy1}. 
The phase profile of such waves resembles a chessboard pattern with lines of $\pi$-phase jumps between neighboring sites.

As a consequence of the anisotropy of axial photorefractive crystals
like strontium barium niobate (SBN), one can distinguish between linear and nonlinear  material response in order to create the desired photonic lattices. An ordinarily polarized light beam only experiences a negligible nonlinearity due to the small electrooptic coefficient and therefore propagates in an almost linear regime.
On the other hand, an extraordinarily polarized light beam is influenced
by a strong photorefractive nonlinearity and propagates in the
nonlinear regime. This includes an anisotropic self focussing of the
incident beam and results in the typical elliptical beam shape at the
output of the photorefractive crystal. 
\begin{figure}[htbp]
  \centering
  \includegraphics[scale=0.8]{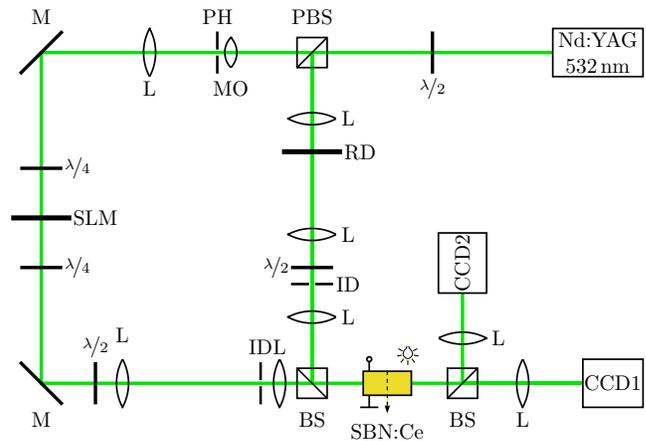}
  \caption{Experimental setup,  (P)BS: (Polarizing) beam splitter, MO: Microscope objective, PH: Pinhole, L: Lens, M: Mirror SLM: Spatial light modulator, ID: Iris diaphragm, RD: Rotating diffuser, CCD1: Real Space Camera, CCD2: Fourier Camera
  \label{setup}}
  \end{figure}

To analyze the structure of the induced refractive index change, two different methods can be used. The first method is given by observing the waveguiding properties of the lattice \cite{Desy1, Denis, Desy2}. As the incident light pattern induces a periodic array of waveguides in the medium, illuminating  the lattice with a broad plane wave leads to guiding of this wave and the output intensity pattern of the guided wave qualitatively maps the induced refractive index change.

Another powerful diagnostic tool for photonic lattices is offered by
the recently demonstrated Brillouin zone spectroscopy
\cite{Bartal}. The aim of this method is a direct visualization of the
lattice structure in Fourier space by mapping the boundaries of the
extended Brillouin zones, which are defined through the
Bragg-reflection planes. To map out the momentum space, the lattice is
probed with a partially spatially incoherent beam, which has a uniform
spatial power spectrum extending over several Brillouin zones and is
broad enough in real space to cover numerous lattice sites. The light
exiting the lattice is then analysed by performing an optical Fourier
transformation and measuring the power spectrum, which contains dark lines
at the borders of the extended Brillouin zones. 

In this contribution we apply both methods to analyze the induced refractive index change of two-dimensional photonic lattices in a photorefractive crystal. For the first time to our knowledge, a Fourier space analysis of two-dimensional nonlinear self-trapped photonic lattices with periodic phase modulation is performed by mapping the borders of the extended Brillouin zones. We compare the results obtained in Fourier space and in real space and demonstrate that both methods are able to show the differences between linear and nonlinear lattices. 
We also show that both methods reveal the orientational dependent structure of the nonlinear induced refractive index change.
\section{Experimental arrangements}
The experimental setup is shown schematically in Fig. \ref{setup}. A
beam derived from a frequency doubled Nd:YAG laser at a wavelength of 532\,nm
is split into two beams by a combination of a half wave plate and a
polarizing beam splitter. The transmitted beam is sent through a
combination of two quarter wave plates and a programmable spatial light
modulator in order to induce the desired pure phase modulation onto the
beam. The modulated beam is then imaged by a high numerical aperture
telescope at the input face of a 20\,mm long
$\mathrm{Sr}_{0.60}\mathrm{Ba}_{0.40}\mathrm{Nb}_2\mathrm{O}_6$
(SBN:Ce) crystal, which is
biased by an externally applied electric field and uniformly
illuminated with a white-light source to control the dark
irradiance. The polarization can be adjusted using a half wave plate
in front of the telescope so that linear and nonlinear lattices can be
induced. By switching off the modulator, the lattice can be illuminated with a broad plane wave to observe the waveguiding properties. 
The second beam is
passed through a rotating diffuser and the partially spatially
incoherent output of the diffuser is imaged at the front face of the
crystal. To ensure that the light will experience the refractive index
modulation of the lattice, the beam is extraordinarily polarized using a
half wave plate.
The output of the crystal is analysed with two CCD cameras. CCD1
monitors the real space output, whereas CCD2 is placed in the focal plane
of a lens to visualize the
Fourier power spectrum of the light exiting the lattice.

In the following experiments, the lattice period is about 22\,$\mathrm{\mu}$m and the applied
electric field is 1\,kV/cm. 
\section{Square Pattern}
\begin{figure}[htb]
 \includegraphics[scale=1]{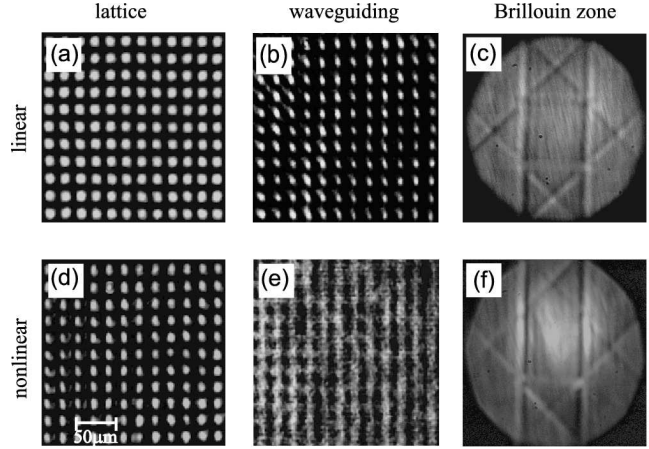}
  \caption{Experimentally generated light intensity patterns for linear and nonlinear square lattices: (a)/(d) lattice output, (b)/(e) guided wave, (c)/(f) Brillouin zone\label{Square}}
\end{figure}
Fig. \ref{Square}(a) depicts the output intensity distribution for
the linear, i.e. ordinarily polarized square lattice at the back face of the crystal. Compared to the nonlinear, i.e extraordinarily polarized lattice output [Fig. \ref{Square}(d)], the only difference is a self focussing of each spot in the nonlinear case.
 The difference between linear and nonlinear square lattices becomes
 more obvious by observing  the waveguiding properties  of the induced lattice [Fig. \ref{Square}(b) and (e)].
The linear wave induces a truly two-dimensional array of effective
focusing lenses and the output intensity distribution of the guided
wave shows a simple square pattern, too [Fig. \ref{Square}(b)].
In contrast to that, the lattice spots in the nonlinear case tend to
fuse into vertical lines due to nonlocality and anisotropy of the
photorefractive response. Therefore, the output
intensity distribution of the guided wave reveals an effectively
one-dimensional refractive index change that basically consists of
vertical lines [Fig. \ref{Square}(e)]. This is in good agreement with previously performed numerical simulations \cite{Desy1,Desy2}.
The Fourier space analysis of the induced refractive index change
confirms this difference between linear and nonlinear lattices. The
Brillouin zone of the linear induced lattice reveals a fully
two-dimensional structure and clearly shows the corresponding first
and second Brillouin zone [Fig. \ref{Square}(c)]. The slightly different contrast for horizontal and vertical lines is due to the anisotropy of the photorefractive material.
On the other hand, the Brillouin zone picture of the nonlinear lattice
is dominated by two vertical lines representing the borders of the first Brillouin zone of the
corresponding one-dimensional stripe pattern [Fig. \ref{Square}(f)]. Additional lines from the originally two-dimensional lattice are also visible, but the two vertical lines are obviously more intense.
 
\section{Diamond pattern}
\begin{figure}[htb]
\includegraphics[scale=1]{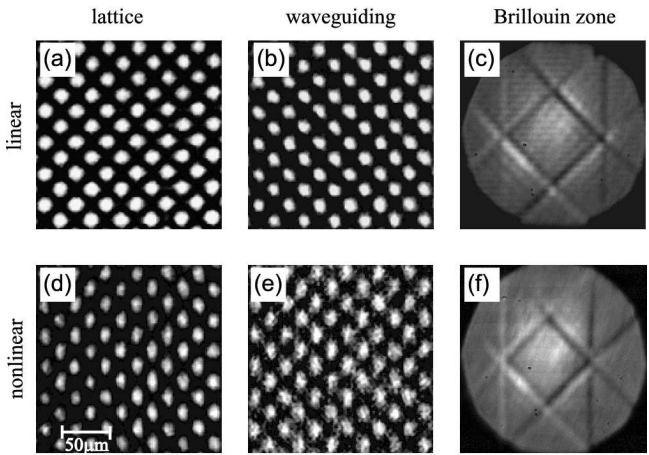}
 \caption{Experimentally generated light intensity patterns for linear and nonlinear diamond lattices: (a)/(d) lattice output, (b)/(e) guided wave, (c)/(f) Brillouin zone  \label{Diamond}}
\end{figure}
Fig. \ref{Diamond} shows the results obtained for the diamond
pattern. Comparing the output intensity distributions of the lattice
waves, the only difference is again found in a self focussing of the
single spots for the nonlinear lattice [Fig. \ref{Diamond}(d)].
As before, the intensity distribution of the guided wave in the linear
lattice basically follows the intensity distribution of the lattice output.
However, the waveguiding of the
nonlinear induced lattice also shows a fully two-dimensional structure
consisting of well separated spots that do not fuse into vertical
lines, as seen for the square pattern [Fig. \ref{Diamond}(e)]. This orientational dependent
structure of the nonlinear induced refractive index change fits well
to our numerical simulations, too \cite{Desy1,Desy2}.
As the waveguiding method does not show any significant difference between the induced refractive index structure for linear and nonlinear lattices, the Fourier space analysis should consequently result in two similar Brillouin zone pictures. This is shown in Fig. \ref{Diamond}(c) and (f). In contrast to the square pattern, the Brillouin zone picture of both, the linear and the nonlinear diamond pattern clearly reveals a  fully two-dimensional structure of the induced refractive index change. 

\section{Conclusions}
We have  generated experimentally linear and nonlinear two-dimensional
self-trapped photonic lattices with different orientations in an anisotropic photorefractive medium.
The differences of the induced refractive index change depending on
polarization and spatial orientation of the periodic lattice wave have
been investigated with two different approaches. For the first time to our knowledge,
partially spatially multi-band excitation has been used to analyze the
structure of two-dimensional nonlinear self-trapped photonic lattices in Fourier
space. Furthermore the waveguiding properies of the lattice have been
observed to achieve a real space analysis of the induced refractive
index change. Both methods clearly reveal the discussed
 structural differences
of the induced refractive index change and  therefore provide powerful diagnostic tools for
further experiments on nonlinear photonic lattices. 

\begin{acknowledgments}
We thank Guy Bartal and Mordechai Segev, Solid State Institute, Technion, Haifa for fruitful discussions on the Brillouin zone spectroscopy. We are also grateful to Zhigang Chen, Department of Physics and Astronomy, San Francisco State University for critical reading of the manuscript. DT acknowledges support by the Konrad-Adenauer-Stiftung e.V.
\end{acknowledgments}

\end{document}